\documentclass[journal=jacsat, manuscript=article, layout=twocolumn]{achemso}
\setkeys{acs}{maxauthors = 10, etalmode=truncate}
\setkeys{acs}{keywords = true}

\usepackage[version=3]{mhchem}
\usepackage[T1]{fontenc}
\usepackage{multicol}
\usepackage{multirow}
\usepackage{mathrsfs}
\usepackage{amsmath}
\usepackage{relsize}
\usepackage{siunitx}
\usepackage{hyperref}
\usepackage{float}
\usepackage{cuted}
\usepackage{threeparttable}
\usepackage{pdfpages}
\usepackage{comment}

\makeatletter
\let\l@addto@macro\relax
\makeatother
\usepackage[fontsize=11pt]{scrextend}

\author{Thomas Possmayer}
\affiliation[LMU]
{Chair in Hybrid Nanosystems, Faculty of Physics, Ludwig-Maximilians-Universität München, 80539 München, Germany}

\author{Allison R. Pessoa}
\affiliation[IFPE]
{Federal Institute of Education, Science and Technology of Pernambuco, 50740-545 Recife-PE, Brazil}
\alsoaffiliation[UFPE]
{Department of Physics, Universidade Federal de Pernambuco, 50740-540 Recife-PE, Brazil}
\email{allison.pessoa@ufpe.br}

\author{Jefferson A. O. Galindo}
\affiliation[UFPE]
{Department of Physics, Universidade Federal de Pernambuco, 50740-540 Recife-PE, Brazil}

\author{Luiz F. dos Santos}
\affiliation[USP]
{Department of Chemistry, Center of Nanotechnology and Tissue Engineering- Mater Lumen Laboratory, Faculty of Philosophy, Science and Letters of Ribeirão Preto, University of São Paulo, 14040-901 Ribeirão Preto-SP, Brazil.}

\author{Rogéria R. Gonçalves}
\affiliation[USP]
{Department of Chemistry, Center of Nanotechnology and Tissue Engineering- Mater Lumen Laboratory, Faculty of Philosophy, Science and Letters of Ribeirão Preto, University of São Paulo, 14040-901 Ribeirão Preto-SP, Brazil.}

\author{Anderson M. Amaral}
\affiliation[UFPE]
{Department of Physics, Universidade Federal de Pernambuco, 50740-540 Recife-PE, Brazil}

\author{Leonardo de S. Menezes}
\affiliation[LMU]
{Chair in Hybrid Nanosystems, Faculty of Physics, Ludwig-Maximilians-Universität München, 80539 München, Germany}
\alsoaffiliation[UFPE]
{Department of Physics, Universidade Federal de Pernambuco, 50740-540 Recife-PE, Brazil}

\title{Boltzmann Thermometry at Cryogenic Temperatures Exploiting Stark Sublevels in Er$^{3+}$/Yb$^{3+}$-Codoped Yttrium Oxide Nanoparticles}

\keywords{Lanthanide ions; Stark sublevels; Accurate Boltzmann thermometry; Cryogenic optical thermometers 
\vspace{0.5cm}}

\let\oldmaketitle\maketitle
\let\maketitle\relax

\begin{document}
\twocolumn[
\begin{@twocolumnfalse}
\oldmaketitle

\begin{abstract}

The development of reliable luminescent nanothermometers for cryogenic applications is essential for advancing quantum technologies, superconducting systems, and other fields that require precise, high-spatial-resolution temperature monitoring. Lanthanide-doped systems are vastly employed to this purpose, and typically perform optimally at or above room temperature when manifold-to-manifold transitions are used. In this work we exploit individual Stark sublevels to demonstrate an optical thermometer based on Er$^{3+}$/Yb$^{3+}$ codoped yttria (Y$_2$O$_3$) nanoparticles that operates effectively across the temperature range from \qtyrange{25}{175}{\kelvin}. This is achieved due to the pronounced crystal field environment of the the Y$_2$O$_3$ host matrix, leading to well-separated Stark lines in the luminescence spectrum of the Er$^{3+}$ ions. By applying the Luminescence Intensity Ratio (LIR) method to transitions originating from two Stark components of the $^4$S$_{3/2}$ manifold of the Er$^{3+}$ ions, we achieve thermal sensitivities up to \qty{1.25}{\percent\per\kelvin} at \qty{100}{K} and temperature resolutions reaching \qty{0.2}{\kelvin}. Our results further experimentally confirm recently published theoretical predictions, demonstrating that thermometric performance is not directly dependent on the average (barycenter) difference of the involved electronic energy levels when using individual Stark transitions to evaluate the LIR. The proposed procedure gives an energy gap calibration that matches the one determined by sample spectroscopy for non-overlapping lines in the luminescence spectrum. These insights provide a robust foundation for the design of high-performance cryogenic thermometers based on rare-earth-doped materials.

\end{abstract}
\end{@twocolumnfalse}
]

\section{Introduction}

The emergence of quantum technologies and the growing demand for precise cryogenic temperature control (at temperature ranges below \qty{140}{\kelvin}) have intensified the need for reliable nanometer- to submicrometer-sized cryothermometers. These sensors are applicable in areas that go beyond quantum computing \cite{Brandl_2016}, reaching the aerospace industry \cite{Kale_2017}, superconductivity-based devices \cite{Steven_2001}, and even medicine and cryobiology \cite{Jain_2021}. Lanthanide ion (Ln$^{3+}$)-doped nanoparticles offer a promising platform for nanoscale and non-invasive optical temperature measurements, showing potential to achieve thermal resolutions below \qty{0.1}{\kelvin} \cite{Brites_2016}. 

Luminescence thermometry based on such nanoparticles has been extensively studied, particularly within the biological temperature range (\qty{0}{\degreeCelsius} to \qty{50}{\degreeCelsius}) \cite{Bednarkiewicz_2021, Goncalves_2021}. The conventional approach relies on the Luminescence Intensity Ratio (LIR) technique involving two thermally coupled (TC) spin-orbit energy manifolds of the Ln$^{3+}$ ions. This method exploits the Boltzmann distribution governing the electronic population of the ions' energy levels \cite{Brites_2016}, which leads to a temperature-dependent emission ratio. Thermometric characteristics such as thermal sensitivity and resolution are directly connected to the energy difference between the TC manifolds \cite{Brites_2016}. For instance, Suta \& Meijerink showed that the most responsively detected temperature is $T_\text{opt} = \Delta E/(2k_\text{B})$, where $\Delta E$ is the manifold energy separation and $k_\text{B}$ is Boltzmann's constant \cite{Suta_Meijerink_2020}. More recently, Pessoa \textit{et al.} have shown theoretically that this energy separation should be treated as an effective value, $\Delta E_\text{eff}$, which also incorporates the oscillator strengths of the electronic transitions involved \cite{Pessoa_2025}.

For most Ln$^{3+}$-doped materials, the effective energy difference between spin-orbit manifolds relevant for thermometry is on the order of \qty{e3}{\per\centi\meter} \cite{Malta_2003, Hanninen_2010}. Specifically, for Er$^{3+}$-based sensors, where the $^2$H$_{11/2}$ and $^4$S$_{3/2}$ manifolds form a TC pair around room temperature, this separation typically ranges from \qty{650}{\per\centi\meter} to \qty{850}{\per\centi\meter}, depending on the host matrix and experimental conditions \cite{Pessoa_2025}. As a result, Boltzmann-type thermometers based on these spin-orbit manifolds operate optimally in the \qtyrange{273}{600}{\kelvin} range \cite{Suta_Meijerink_2020}.

To meet the growing need for cryogenic-range luminescent thermometers, several studies have proposed alternative approaches that exploit the temperature-dependent luminescent behavior of non-TC levels. These include mechanisms such as temperature-dependent non-radiative energy transfer \cite{Wei_2025, Ren_2015, Louika_2017}, phonon-assisted energy migration \cite{Xue_2015}, and variations in luminescence decay time \cite{Bolek_2021}. Only recently, some works have turned toward exploiting individual Stark sublevels within the TC manifolds \cite{Skul_2025, Yanru_2024, Dodson_2024, Boldyrev_2024, Zhang_2023, Alexey_2023, Kitos_2020, Shang_2019, Yu2021}. These Stark sublevels arise from the splitting of otherwise degenerate spin-orbit levels due to local electric fields in the crystal environment \cite{Malta_2003}. With typical separations on the order of \qty{e2}{\per\centi\meter}, they enable optimal thermal responses around 40 K \cite{Suta_Meijerink_2020}. However, the applicability of this approach depends strongly on the host matrix, as it must induce sufficiently large and spectrally resolvable Stark splittings. Yttria (Y$_2$O$_3$) is  particularly well-suited for this purpose, as it produces pronounced Stark splitting and narrow emission lines from embedded Ln$^{3+}$ ions \cite{Pessoa_2023, Casabone_2018}.

While optical cryothermometers based on Stark sublevels have been proposed, a comprehensive understanding of the factors governing their sensitivity and accuracy remains unavailable. A recent theoretical framework has exposed and discussed some of these problems \cite{Pessoa_2025}. Specifically, it has been shown that the energy difference between the barycenters of two given Stark lines does not directly determine thermal characteristics, such as the thermometer's relative sensitivity and accuracy, as they do not represent the real relative population distribution and energy difference between the TC Stark sublevels \cite{Pessoa_2025}.

In this work, we leverage this insight to choose adequate lines to avoid spectral overlapping and intruding bands - both of which are known to degrade the accuracy of ratiometric thermometry, as already demonstrated for manifold-to-manifold transitions \cite{Pessoa_2023}. We demonstrate a Boltzmann-type optical thermometer operating in the cryogenic range between \qty{25}{\kelvin} and \qty{175}{\kelvin} based on selected Stark-to-Stark transitions from the Er$^{3+}$ $^4$S$_{3/2}$ manifold to the ground state ($^4$I$_{15/2}$). Our results validate recently published theoretical predictions\cite{Pessoa_2025} and provide a step forward to achieve accurate and high-precision cryogenic thermometry using rare-earth-doped systems.

\begin{figure*}[h!] 
\begin{center}
	\includegraphics[width=\linewidth]{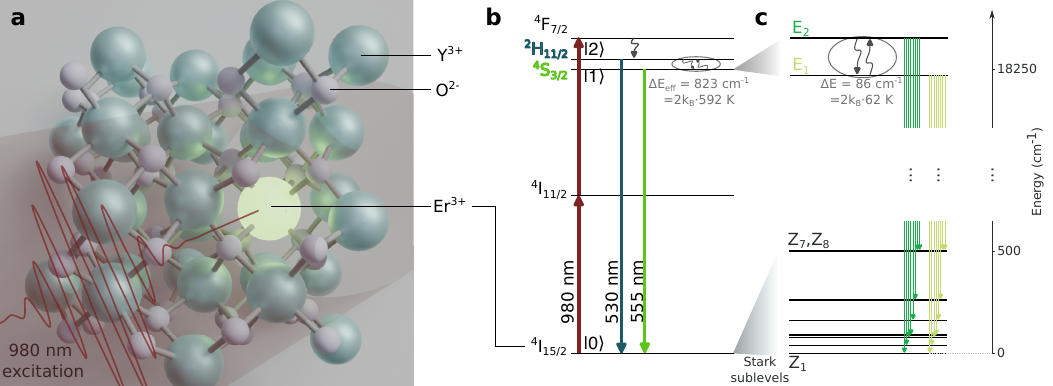}
	\caption{\textbf{a}, Unit cell of Y$_2$O$_3$ with a doping Er$^{3+}$ ion (bright green) occupying a $C_2$ symmetry site, emitting upconverted green light under excitation at a wavelength of \qty{980}{nm}. Turquoise (grey) spheres represent yttrium (oxygen) atoms. \textbf{b}, Simplified energy-level diagram of the Er$^{3+}$ ions, showing the relevant levels for nanothermometry via two-photon optical excitation. Upward straight arrows show possible energy transfer mechanisms from Yb$^{3+}$ ions (omitted for clarity) or direct absorption from Er$^{3+}$ ions. Downward straight arrows represent radiative decays; curly arrows indicate non-radiative relaxation processes. Circled levels are considered TC around room temperature and above. \textbf{c}, Stark sublevels of the corresponding spin-orbit manifolds. The circled levels are candidates for being used as TC levels in LIR-based optical thermometry experiments performed at cryogenic temperatures.}
	\label{fig:experimental}
\end{center}
\end{figure*}

\section{Results and Discussions}

\subsection{Upconversion Spectrum at Low Temperatures}

In thermometry experiments employing Yb$^{3+}$ / Er$^{3+}$ codoped systems, it is common to excite the TC levels via a two-photon upconversion scheme: Illumination at \qty{980}{\nano\meter} initially populates the $^4$F$_{7/2}$ manifold of Er$^{3+}$ ions through a two-step upconversion process assisted by the Yb$^{3+}$ ions in a well-known mechanism of donor-acceptor energy transfer \cite{Hinojosa_2003}. Fig. \ref{fig:experimental}a depicts the unit cell of the used yttria matrix with an Er$^{3+}$ ion substituting an Y$^{3+}$ ion and Fig. \ref{fig:experimental}b shows the relevant photophysical processes under the considered excitation scheme (not showing the Yb$^{3+}$ ions' energy levels for simplicity).

Following the excitation, non-radiative relaxation via electron-phonon interactions transfers population from $^4$F$_{7/2}$ to the metastable $^2$H$_{11/2}$ and $^4$S$_{3/2}$ manifolds. These two levels are considered TC at room temperature, as their energy separation is small enough to allow thermalization in a timescale of nanoseconds to hundreds of nanoseconds, depending on the host matrix \cite{Pessoa_2025}, leading to a population probability distribution well-described by Boltzmann statistics. Both levels subsequently decay radiatively to the ground state ($^4$I$_{15/2}$), emitting photons in the green spectral range. The precise emission wavelengths are determined by the Stark sublevel structure of the excited and ground manifolds (Fig. \ref{fig:experimental}c).

The energy separation and remaining degeneracies of the resulting Stark sublevels are mainly governed by the host matrix. In the Y$_2$O$_3$ crystalline material, Er$^{3+}$ and Yb$^{3+}$ ions occupy sites with $C_2$ and $C_{3i}$ point symmetry, with the optical emission in the green spectral region predominantly stemming from the former site \cite{Kisliuk_1964}. Under this local symmetry, the $^2$H$_{11/2}$ and $^4$S$_{3/2}$ manifolds split into six and two Stark sublevels, respectively \cite{Hanninen_2010}. The latter thus provides a potential further pair of TC levels at much lower temperatures (Fig. \ref{fig:experimental}c). Similarly, the ground state $^4$I$_{15/2}$ splits into eight sublevels. Each of these Stark sublevels is a Kramer's doublet and therefore doubly degenerate. \cite{Hanninen_2010} 

Assuming that their electronic population follows Boltzmann statistics, the occupation probability of a Stark sublevel $|i\Gamma_{k}\rangle$ within the manifold $|i\rangle$ is given by \cite{Pessoa_2025}

\begin{equation}
	p_{ik}(T) = \frac{g_{ik} \exp{\left(-\frac{E_{ik}-E_{0}}{k_{\text{B}}T}\right)}}{\mathlarger{\sum}_{j}\mathlarger{\sum}_{l=1}^{L_j} g_{jl}\exp{\left(-\frac{E_{jl}-E_{0}}{k_{\text{B}}T}\right)}} \; ,
	\label{eq:boltzmann_distribution}
\end{equation}

\begin{figure*}[h!] 
\begin{center}
	\includegraphics[width=\linewidth]{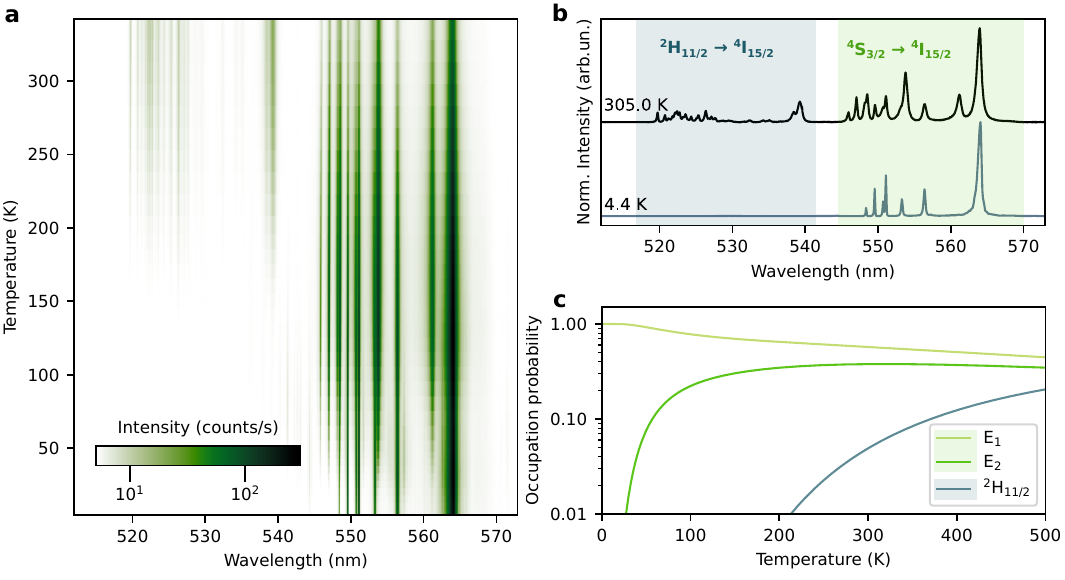}
	\caption{\textbf{a}, Temperature-dependent spectra of the upconverted luminescence in the green spectral region emitted by Y$_2$O$_3$: Yb$^{3+}$ / Er$^{3+}$ nanoparticles under \qty{980}{nm} excitation. Emission from the $^2$H$_{11/2}$ manifold is negligible at cryogenic temperatures. \textbf{b}, Normalized luminescence spectra at \qty{4.4}{\kelvin} and \qty{305}{\kelvin} (taken from panel \textbf{a}) with highlighted manifold transitions. \textbf{c}, Calculated occupation probability distribution of the thermally coupled levels of Er$^{3+}$. Labels E$_1$ and E$_2$ correspond to the two Stark components of the $^4$S$_{3/2}$ manifold; $^2$H$_{11/2}$ denotes the combined population of all Stark sublevels of the $^2$H$_{11/2}$ manifold.}
	\label{fig:spectrum_manifold}
\end{center}
\end{figure*}

\noindent where $g_{jl}$ is the degeneracy of the Stark sublevel $|j\Gamma_{l}\rangle$ and $E_{jl}$ is its energy. The reference energy $E_{0}$ corresponds to the lowest Stark sublevel of the $^4$S$_{3/2}$ manifold. $k_{\text{B}}$ is Boltzmann's constant and $T$ is the absolute temperature. The sum over $j$ includes both TC manifolds, while the sum over $l$ runs over all $L_j$ Stark sublevels within each manifold. 

Temperature variation from cryogenic to room temperature therefore induces significant changes in the electronic population, and consequently also in the emission spectrum (Fig. \ref{fig:spectrum_manifold}a). These changes arise both from thermal redistribution between the $^2$H$_{11/2}$ and $^4$S$_{3/2}$ manifolds, and from population changes among their individual Stark sublevels. Fig. \ref{fig:spectrum_manifold}b shows the measured luminescence spectra at \qty{4.4}{\kelvin} and \qty{305}{\kelvin}, highlighting the higher population probability of the lower-energy manifold at low temperatures. The observed spectral lines have average width in the order of \qty{0.8}{\nano\meter}. Emission lines with wavelengths between \qty{517}{\nano\meter} and \qty{542}{\nano\meter} correspond to transitions from Stark sublevels of $^2$H$_{11/2}$ to Stark sublevels of the ground state $^4$I$_{15/2}$ (respecting possible selection rules), while those between \qty{545}{\nano\meter} and \qty{570}{\nano\meter} correspond to transition between the Stark sublevels of $^4$S$_{3/2}$ $\rightarrow$ $^4$I$_{15/2}$.

To further illustrate the temperature-dependent population in the sublevels, Fig. \ref{fig:spectrum_manifold}c shows calculated occupation probabilities for the two Stark sublevels of the $^4$S$_{3/2}$ manifold, labeled as E$_1$ and E$_2$ in Fig. \ref{fig:experimental}b (following the empirical notation), as well as the combined occupation probability of all sublevels in the $^2$H$_{11/2}$ manifold (labeled as H). These calculations were performed using the levels' energies provided by Kisliuk \textit{et al.} \cite{Kisliuk_1964} for Er$^{3+}$ in a Y$_2$O$_3$ matrix. At room temperature, the $^2$H$_{11/2}$ manifold becomes appreciably populated, while E$_1$ and E$_2$ are close to being equally populated due to their small energy separation. At cryogenic temperatures, however, occupation of the $^2$H$_{11/2}$ manifold becomes negligible, and the higher-lying E$_2$ sublevel is depopulated in favor of E$_1$.

\subsection{Identifying Stark Sublevels}

\begin{figure}[h!] 
\begin{center}
	\includegraphics[width=\linewidth]{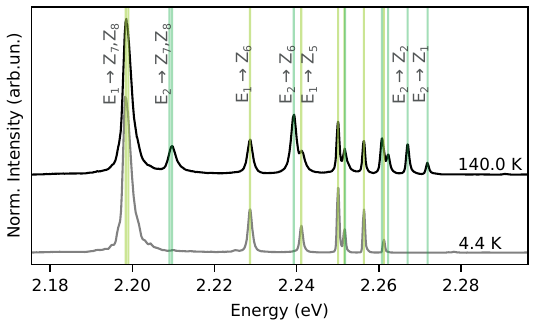}
	\caption{Normalized luminescence spectra of the $^4$S$_{3/2}$ $\rightarrow$ $^4$I$_{15/2}$ band, plotted in energy scale, at \qty{4.4}{\kelvin} and \qty{140}{\kelvin} (offset for clarity). Selected transitions relevant to the thermometric analysis are labeled using empirical notation.}
	\label{fig:spectrum_stark}
\end{center}
\end{figure}

To investigate the population distribution of the TC manifolds by means of the Boltzmann law, we spectroscopically measured the energy separation between their Stark sublevels by analysing the luminescence spectrum. This requires transforming the spectral curve ($I(\lambda)$) to the energy domain, using \cite{Suta_Meijerink_2020}

\begin{equation}
\tilde I(E) = \left(\frac{\lambda^2}{hc}\right)I(\lambda) \,.
\label{eq:energy_wvl_transf}
\end{equation}

\noindent where $h$ is Planck's constant and $c$ the speed of light in vacuum. Fig. \ref{fig:spectrum_stark} shows the $^4$S$_{3/2}$ $\rightarrow$ $^4$I$_{15/2}$ luminescence band converted to the energy scale, along with the assignments of selected Stark-Stark transitions for two different temperatures (a broader temperature sweep is presented in Fig. S1 of the Supporting Information). A similar analysis was performed for the $^2$H$_{11/2}$ $\rightarrow$ $^4$I$_{15/2}$ luminescence band at temperatures above 200 K. The extracted energies of the Stark sublevels - obtained at \qty{140}{\kelvin} for $^4$S$_{3/2}$ and at \qty{260}{\kelvin} for $^2$H$_{11/2}$ - are listed in table \ref{tab:energy_levels}.

The energy splitting between the two Stark components of the $^4$S$_{3/2}$ manifold was experimentally determined to be \qty{86}{\per\centi\meter} from the emission spectra. Therefore, at \qty{4.4}{\kelvin} (up to approximately $\sim$\qty{20.0}{\kelvin}),  nearly all excited ions reside in the lowest Stark sublevel of $^4$S$_{3/2}$ (E$_1$), as predicted in Fig. \ref{fig:spectrum_manifold}c. Accordingly, the luminescence spectrum at this temperature in Fig. \ref{fig:spectrum_stark} exhibits only eight distinct peaks - two of which are spectrally overlapped - corresponding to radiative decays from E$_1$ to the eight Stark components of the ground state $^4$I$_{15/2}$ (Z$_1$ to Z$_8$). 

As the temperature increases and E$_2$ becomes thermally populated, eight additional spectral lines emerge in the $\qty{540}{\nano\meter} - \qty{570}{\nano\meter}$ range. As shown in Fig. \ref{fig:spectrum_manifold}a, above \qty{200}{\kelvin} the $^2$H$_{11/2}$ $\rightarrow$ $^4$I$_{15/2}$ emission band also appears, while the intensity of the $^4$S$_{3/2}$ $\rightarrow$ $^4$I$_{15/2}$ transition begins to decrease.

Interestingly, we observed that the overall emission intensity of the $^4$S$_{3/2}$ $\rightarrow$ $^4$I$_{15/2}$ transition reaches a maximum near \qty{140}{\kelvin} (see Fig. S2 of the SI). This behavior is likely attributable to mechanical shifts in the optical illumination or collection paths with varying temperature. Additionally, the number of excited ions depends on the efficiency of the upconversion process, which may itself be temperature dependent due to the thermal activation of energy transfer rates \cite{Yu_2014}. However, as ratiometric thermometry involves comparing emission intensities at the same temperature, such variations do not compromise the accuracy or precision of the thermometer.

\begin{table}[h]
\caption{\label{tab:energy_levels}%
Measured energies of the Stark sublevels within the relevant Er$^{3+}$ spin-orbit manifolds in Y$_2$O$_3$ at \qty{140}{K} and \qty{260}{K} for the $^4$S$_{3/2}$ and $^2$H$_{11/2}$ manifold, respectively. Values are compared to literature data from Kisliuk \textit{et al.} \cite{Kisliuk_1964}.
}
\resizebox{\columnwidth}{!}{%
\centering
\begin{tabular}{cccc}
\hline
\begin{tabular}[c]{@{}c@{}}Spin-orbit \\ manifold\end{tabular} & \begin{tabular}[c]{@{}c@{}}Empirical \\ notation\end{tabular} & \begin{tabular}[c]{@{}c@{}}Energy (cm$^{-1}$)\\ (this work)\end{tabular} & \begin{tabular}[c]{@{}c@{}}Energy (cm$^{-1}$)\\ (Ref [\!\!\citenum{Kisliuk_1964}])\end{tabular} \\ \hline
 & Z$_1$ & 0 & 0 \\
 & Z$_2$ & 39 & 39 \\
 & Z$_3$ & 77 & 76 \\
\multirow{2}{*}{$^4$I$_{15/2}$} & Z$_4$ & 90 & 89 \\
 & Z$_5$ & 162 & 158 \\
 & Z$_6$ & 263 & 258 \\
 & Z$_7$ & 502 & 500 \\
 & Z$_8$ & 507 & 500 \\
 &  &  &  \\
\multirow{2}{*}{$^4$S$_{3/2}$} & E$_1$ & 18238 & 18231 \\
 & E$_2$ & 18324 & 18318 \\
 &  &  &  \\
 & F$_1$ & 19041 & 19038 \\
 & F$_2$ & 19050 & 19045 \\
\multirow{2}{*}{$^2$H$_{11/2}$} & F$_3$ & 19077 & 19072 \\
 & F$_4$ & 19192 & 19187 \\
 & F$_5$ & 19223 & 19218 \\
 & F$_6$ & 19247 & 19243 \\ \hline
\end{tabular}}
\end{table}

\subsection{Boltzmann Thermometry using Stark Sublevels}

For the relative population of two TC levels to be correctly described by a Boltzmann distribution, it is required that the radiative decay rates of the TC levels are much smaller than the non-radiative, phonon-assisted thermalization rate. This condition is satisfied in commonly used host matrices, since the electron-phonon interaction rate leading to thermalization is typically five orders of magnitude greater than the radiative decay rate \cite{Dechao_2016}. However, photophysical processes such as cross-relaxation, excited-state absorption, or surface quenching - if occurring at rates comparable to that of phonon-mediated relaxation - can disrupt the Boltzmann equilibrium \cite{Suta_Miroslav_2020, Galvao2_2021}. It is therefore essential to ensure that experimental conditions allow for valid Boltzmann statistics.

The ratiometric Boltzmann method relies on measuring the ratio of integrated intensities of two emission bands originating from thermally coupled manifolds. In standard implementations, the LIR is calculated by integrating the entire manifold-to-manifold emission bands. However, if the host matrix allows for spectral resolution of individual Stark transitions - as is the case in Y$_2$O$_3$: Yb$^{3+}$/Er$^{3+}$ - then specific Stark-Stark lines can be spectrally isolated and used to compute the LIR and consequently to accurately determine the system's temperature.

The intensity of a spectral line is proportional to the total photon emission rate and can be obtained by integrating the luminescence signal ($I(\lambda)$) over the relevant wavelength range \cite{Suta_Meijerink_2020}. Care must be taken to avoid the contribution of accidentally superimposed luminescence bands which are not related to the relevant TC levels, contributing to the calculation of the LIR and leading to inaccuracies in the temperature readout\cite{Pessoa_2023}. If such intruding bands are identified, they can be separated a posteriori through the use of non-arbitrary methods\cite{Galindo_2023} or alternative excitation strategies\cite{Pessoa_2024}.

For a transition from a Stark sublevel $|i\Gamma_k\rangle$ (with $i \in \{^4\text{S}_{3/2},\, ^2\text{H}_{11/2}\}$ in our case) to a ground-state sublevel $|^4\text{I}_{15/2}\,\Gamma_m\rangle \equiv |0\Gamma_m\rangle$, the integrated intensity over a wavelength interval $[\lambda_1,\, \lambda_2]$ without overlapping spectral lines is proportional to the excited energy level's population $n_{ik}$ via \cite{Pessoa_2025}

\begin{equation}
\int_{\lambda_1}^{\lambda_2} I(\lambda)\, \text{d}\lambda = \eta(\tilde\lambda_{12})\Phi_{ik,0m} = \eta(\tilde\lambda_{12}) \,A_{ik,0m}  \, n_{ik} \quad ,
	\label{eq:photons_emitted}
\end{equation}

\noindent where $\eta(\tilde\lambda_{12})$ is the average detection efficiency across the chosen interval, incorporating system optics and the detector sensitivity. $\Phi_{ik,0m}$ is the photon emission rate in the transition. It is equal to the population in the excited Stark sublevel $|i\Gamma_k\rangle$, $n_{ik}$, multiplied by the Einstein coefficient for the radiative transition, $A_{ik,0m}$. Similarly, one can obtain the integrated intensity of a complete manifold-to-manifold transition by summing all Stark-Stark contributions.

Pessoa \textit{et al.} \cite{Pessoa_2025} have shown that when employing the LIR method by integrating the complete manifold-to-manifold transitions, the LIR as a function of temperature ($R(T)$) is a weighted sum of exponentials. This can be approximated by a single exponential with effective parameters:

\begin{equation}
\begin{aligned}
R(T)_\text{manifold} &= \frac{\mathlarger{\sum}_{km}\, A_{Hk,0m} \, g_{Hk} \, \exp{\left(-E_{Hk}/k_{\text{B}}T\right)}}{\mathlarger{\sum}_{l m}\, A_{Sl,0m} \, g_{Sl} \, \exp{\left(-E_{Sl}/k_{\text{B}}T\right)}} \\
&\approx C_{\text{eff}} \cdot \exp{\left(-\frac{\Delta E_{\text{eff}}}{k_{\text{B}}T}\right)}\; ,
\end{aligned}
\label{eq:LIR_maifold_manifold}
\end{equation}

\noindent where $Hk$ and $Sl$ label Stark sublevels of the $^2\text{H}_{11/2}$ and $^4\text{S}_{3/2}$ manifolds, respectively. The parameters $C_\text{eff}$ and $\Delta E_{\text{eff}}$ can be predicted by expanding Eq. \eqref{eq:LIR_maifold_manifold} around a central $\beta_c = (k_\text{B}\, T_c)^{-1}$ (where $T_c$ is the central temperature of the range under consideration) and truncating to first order:

\begin{equation}
\begin{aligned}
&\ln C_\text{eff} = \ln R(T_c) + T_c\cdot S_r(T_c) \\
&\Delta E_\text{eff} = S_r(T_c) \cdot k_\text{B}T_c^2 \quad , 
\label{eq:DeltaE_eff_andC_eff}
\end{aligned}
\end{equation}

\noindent where $S_r(T) = \frac{1}{R}\frac{\partial R}{\partial T}$ is the thermometer's relative sensitivity, a common figure of merit for comparing thermometer performances.

In practical applications, $C_\text{eff}$ and $\Delta E_\text{eff}$ are typically determined through prior calibration, by acquiring a set of emission spectra at externally measured temperatures and fitting $R(T)$ using Eq. \eqref{eq:LIR_maifold_manifold}. However, as shown in Fig. \ref{fig:spectrum_manifold}, the $^2$H$_{11/2}$ manifold is scarcely populated below \qty{200}{\kelvin} (less than 1 \% of the total population), rendering manifold-to-manifold thermometry ineffective in the cryogenic regime due to vanishing sensitivities. In contrast, Stark sublevels with energy separations on the order of \qty{100}{\per\centi\meter} can enable Boltzmann thermometry at these lower temperatures. In this case, the LIR for a Stark-to-Stark transition $|i\Gamma_k\rangle$ to $|j\Gamma_l\rangle$ is given by a single exponential function, through \cite{Pessoa_2025}

\begin{equation}
R(T)_\text{Stark} = \frac{A_{ik,0m}}{A_{jl,0m'}} \frac{g_{ik}}{g_{jl}} \cdot \exp{\left(-\frac{\Delta E_{ik,jl}}{k_{\text{B}}T}\right)} \; ,
\label{eq:LIR_Stark_Stark}
\end{equation}

\noindent where it is possible to have $i=j$ and $k\neq l$, which corresponds to using two distinct Stark sublevels from the same manifold. Here, $\Delta E_{ik,jl}$ is simply the actual energy separation between the TC Stark sublevels, assuming perfect Boltzmann thermalization and accurate temperature readout. 

This approach allows direct extraction of microscopic quantities from $R(T)_\text{Stark}$ through curve fitting. This is not possible for the manifold-to-manifold approach since the expression for $\Delta E_\text{eff}$ (Eq. \eqref{eq:DeltaE_eff_andC_eff}) involves all oscillator strengths between the Stark sublevels of the TC manifolds and the ground state. These Stark-Stark oscillator strengths are not straightforward to calculate since 4f-4f transitions are parity-forbidden, making the use of Judd-Ofelt theory necessary, which requires other specific details about the Ln$^{3+}$-host system\cite{Walsh_2006}. Therefore, knowing $\Delta E_\text{eff}$ does not yield direct information about microscopic parameters.

\subsection{Thermometric Characterization}

According to the results presented in Table \ref{tab:energy_levels}, the Stark sublevels E$_1$ and E$_2$ of the $^4$S$_{3/2}$ manifold are separated by \qty{86}{\per\centi\meter}, while their radiative decay to the ground state can result in spectral lines separated by more than \qty{590}{\per\centi\meter} due to splitting in the ground-state manifold. To calculate the bands' intensities, we separated the Stark lines by fitting them with Voigt profiles, as shown in Fig. S3 of the SI. Despite their differences in spectral separation, fitting the resulting $R_\text{Stark}(T)$ with a Boltzmann factor from Eq. \eqref{eq:LIR_Stark_Stark} consistently yields $\Delta E_{\text{eff}} = \Delta E_{\text{E}_1,\text{E}_2} \approx \; $\qty{86}{\per\centi\meter}, as spectroscopically determined.

\begin{figure}[h!] 
\begin{center}
	\includegraphics[width=\linewidth]{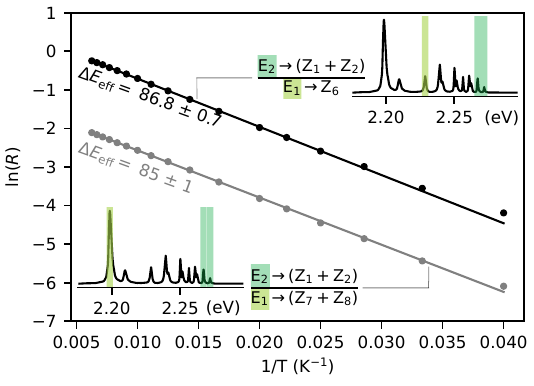}
	\caption{Thermometric characterization using different spectral lines arising from the same thermally coupled Stark sublevels. Insets show spectra from Fig. \ref{fig:spectrum_stark} at \qty{140}{K}, with the transitions used for the LIR evaluation highlighted.}
	\label{fig:thermometry_stark}
\end{center}
\end{figure}

This is illustrated in Fig. \ref{fig:thermometry_stark}, using two LIR pairs:
i) $\text{E}_2 \rightarrow (\text{Z}_1 + \text{Z}_2)$ vs. $\text{E}_1 \rightarrow (\text{Z}_7 + \text{Z}_8)$, which have a spectral separation of \qty{560.6}{\per\centi\meter}; and 
ii)  $\text{E}_2 \rightarrow (\text{Z}_1 + \text{Z}_2)$ vs. $\text{E}_1 \rightarrow \text{Z}_6$, which are separated by \qty{321.3}{\per\centi\meter} (see Fig. \ref{fig:spectrum_stark}b for the line assignments). The linear relationship between $\ln(R)$ and $1/T$ confirms that the population can be described by Boltzmann statistics, while the slope yields the effective energy separation, which is statistically equal to $\qty{86}{cm^{-1}}$ in both cases.

Here it is worth stressing that since the relative sensitivity for Stark-Stark transitions $S_r = {\Delta E_{\text{E}_1,\text{E}_2}}/{(k_{\text{B}}T^2)}$ depends only on the true energy difference between the crystal-field states, it has no relation to the separation between the resulting lines in the spectrum - allowing one to select the most convenient, separable lines.

Another figure of merit characterizing a thermometer is the thermal resolution $\sigma_T$, defined as the uncertainty in the temperature measurement. It is calculated via propagation in uncertainties in $\Delta E_{\text{eff}}$, $C_\text{eff}$ and the measured LIR: \cite{Pessoa_2025}

\begin{equation}
\begin{aligned}
\sigma_T^2 &= \frac{1}{S_r(T_c)^2} \left[ \left(\frac{\sigma_\alpha}{T_c}\right)^2 + \sigma_\beta^2 + \left(\frac{\sigma_R}{R}\right)^2 - 2\frac{\sigma_{\alpha \beta}}{T_c}\right].
\label{eq:thermal_resolution_solved}
\end{aligned}
\end{equation}

\noindent where $\alpha = \Delta E_{\text{eff}}/k_\text{B}$, $\beta = \ln(C_\text{eff})$, and $\sigma_\alpha$, $\sigma_\beta$ and $\sigma_R$ are the uncertainties of their corresponding variables. Note that the uncertainty in determining $\alpha$ and $\beta$ also depends on the uncertainty of the external thermometer used for calibration of the nanothermometer. $\sigma_{\alpha \beta}$ is the covariance between these two variables, and it was previously shown that it can be a relevant correction \cite{Galindo_2021, GalindoCorr_2021}. Table \ref{tab:thermometry_comparison} presents the full thermometric characterization for selected Stark transitions and compares them to the manifold-based case. 

\begin{table*}[h!]
\caption{\label{tab:thermometry_comparison}%
Thermometric characterization using various spectral lines arising from the same thermally coupled Stark sublevels, including their standard deviations (in parentheses). For completeness, the final row presents the corresponding results using complete manifold-to-manifold integration.
}
\bgroup
\def\arraystretch{1.8}

\centering
\begin{tabular}{cccccccc}
\hline
LIR assignment 
& \begin{tabular}[c]{@{}c@{}}Temperature\\ range (K)\end{tabular} 
& $C_\text{eff}$ 
& \begin{tabular}[c]{@{}c@{}}$\Delta E_\text{eff}$ \\ (cm$^{-1}$)\end{tabular} 
& \begin{tabular}[c]{@{}c@{}}$\Delta E_\text{bary}$$^a$ \\ (cm$^{-1}$)\end{tabular} 
& \begin{tabular}[c]{@{}c@{}}$\Delta E_\text{avg}$$^a$ \\ (cm$^{-1}$)\end{tabular}
& \begin{tabular}[c]{@{}c@{}}$S_r (T_c)$$^a$ \\ (\% K$^{-1}$)\end{tabular} 
& \begin{tabular}[c]{@{}c@{}}$\sigma_T $$^a$ \\ (K)\end{tabular}\\
\hline

$\frac{\text{E}_2 \rightarrow (\text{Z}_1 + \text{Z}_2)}{\text{E}_1 \rightarrow \text{Z}_6}$ & 25 - 175 & 1.72(1) & 86.8(7) & 321.3(1) & 86 & 1.25(1) & 0.9\\

$\frac{\text{E}_2 \rightarrow (\text{Z}_1 + \text{Z}_2)}{\text{E}_1 \rightarrow (\text{Z}_7 + \text{Z}_8)}$ & 25 - 175 & 0.260(4) & 85(1) & 560.6(1) & 86 & 1.22(2) & 1.3\\

$\frac{\text{E}_2 \rightarrow (\text{Z}_7 + \text{Z}_8)}{\text{E}_1 \rightarrow (\text{Z}_7 + \text{Z}_8)}$ & 25 - 175 & 0.327(2) & 64.1(4) & 86.7(1) & 86 & 0.922(6) & 0.9\\

$\frac{\text{E}_2 \rightarrow \text{Z}_6}{\text{E}_1 \rightarrow \text{Z}_5}$ & 25 - 175 & 6.90(1) & 70.8(1) & 14.9(1) & 86 & 1.018(2) & 0.2\\

Manifold &  &  &  &  &  & \\ \hline
$\frac{^2\text{H}_{11/2} \, \rightarrow \, ^4\text{I}_{15/2}}{^4\text{S}_{3/2} \, \rightarrow \, ^4\text{I}_{13/2}}$ & 200 - 350 & 12.4(3) & 823(6) & 929(1) & 859 & 1.57(1) & 1.0\\ \hline
\end{tabular}
\egroup

 \begin{tablenotes}
   \item [a] $^a$ At the central temperature of the range, $T_c$.
 \end{tablenotes} 
\end{table*}

These results confirm that Stark-level thermometry is viable in the \qtyrange{25}{175}{\kelvin} range in Y$_2$O$_3$: Yb$^{3+}$/Er$^{3+}$ systems. For instance, the pair ${\text{E}_2 \rightarrow (\text{Z}_1 + \text{Z}_2)}$ and ${\text{E}_1 \rightarrow \text{Z}_6}$ yields high relative sensitivity in the cryogenic regime (comparable to that of using complete manifolds at room temperature), and yields accurate temperatures. Those lines are well-separated from other lines, thereby reducing artifacts in the temperature measurement related to luminescence band overlapping. In contrast, using ${\text{E}_2 \rightarrow (\text{Z}_1 + \text{Z}_2)}$ and ${\text{E}_1 \rightarrow (\text{Z}_7 + \text{Z}_8)}$ introduces partial overlap with $\text{E}_2 \rightarrow (\text{Z}_7 + \text{Z}_8)$, which increases uncertainties at similar sensitivities. Additionally, its lower LIR further increases the thermal uncertainty - consistent with Eq. \eqref{eq:thermal_resolution_solved}.

The other two investigated pairs ($\text{E}_2 \rightarrow (\text{Z}_7 + \text{Z}_8)$ vs. $\text{E}_1 \rightarrow (\text{Z}_7 + \text{Z}_8)$ and $\text{E}_2 \rightarrow \text{Z}_6$ vs $\text{E}_1 \rightarrow \text{Z}_5$) result in higher sensitivities, but feature spectrally superposed lines, which can compromise the accuracy of the extracted values; in particular, the latter pair has a peak energy separation of just \qty{15}{\per\centi\meter}. It also showed the lowest achieved thermal uncertainty of \qty{0.2}{\kelvin} at $T_c = $ \qty{100}{\kelvin}, with $S_r = 1.018 \, \% \, K^{-1}$. This was caused by its higher LIR values, thus improving thermal resolution at the expense of fitting parameters affecting the extracted $\Delta E_\text{eff}$. 

Regarding the manifold-to-manifold characterization, also shown in Table \ref{tab:thermometry_comparison}, we observe that $\Delta E_\text{eff} \neq \Delta E_\text{bary} \neq \Delta E_\text{avg}$, as theoretically predicted \cite{Pessoa_2025}. Notably, the difference between $\Delta E_\text{eff}$ and $\Delta E_\text{bary}$ exceeds \qty{100}{\per\centi\meter}, implying that the use of $\Delta E_\text{bary}$ instead of $\Delta E_\text{eff}$ in LIR thermometry without careful calibration may lead to temperature errors of more than \qty{35}{\kelvin} at room temperature (\qty{295}{K}). This was estimated by Eq. \eqref{eq:thermal_resolution_solved}, where we have considered $\sigma_\alpha = 100/k_B$ and $S_r = 1.36$. The other uncertainties were set to zero to analyse only the influence of $\sigma_\alpha$ in this estimation.

\section{Conclusions}

We have demonstrated that an optical nanothermometer based on Y$_2$O$_3$: Yb$^{3+}$/Er$^{3+}$ nanoparticles can operate effectively in the \qtyrange{25}{175}{\kelvin} temperature range. The luminescence spectrum of these systems exhibits well-resolved Stark lines with full widths at half maximum around \qty{0.8}{\nano\meter} and minimal spectral overlap. This spectral resolution enables the isolation of individual Stark transitions via Voigt profile fitting. By applying the LIR-based method using the two Stark sublevels of the $^4$S$_{3/2}$ spin-orbit manifold of the Er$^{3+}$ ions, we achieved a thermal sensitivity of \qty{1.25}{\percent\per\kelvin}, and a thermal resolution of \qty{0.2}{\kelvin}. The energy difference between these Stark sublevels is \qty{86}{\per\centi\meter}, measured directly through the luminescence spectra. Although their radiative decays to the Stark-split ground state result in emission lines separated by more than \qty{590}{\per\centi\meter}, we have shown that the thermometer’s performance depends solely on the true energy separation of the Stark levels, independent of the energy barycenter of the chosen emission lines. These findings confirm recent theoretical predictions concerning the principles of temperature readout in such systems\cite{Pessoa_2025}.

\section{Methods}
\subsection{Sample Synthesis}
Nanocrystalline Er$^{3+}$ / Yb$^{3+}$ co-doped Y$_2$O$_3$ was synthesized through a homogeneous precipitation method followed by controlled thermal treatment to ensure phase purity and crystallinity. Initially, Er$^{3+}$ and Yb$^{3+}$ co-doped yttrium hydroxycarbonate [Y(OH)CO$_3\cdot$nH$_2$O] was prepared and employed as a precursor. The homogeneous precipitation was achieved through urea thermolysis, conducted in an aqueous solution of yttrium nitrate hexahydrate (Y(NO$_3$)$_3\cdot_6$H$_2$O, 99.8\% purity, Sigma-Aldrich\textsuperscript{\textregistered}) and urea (99.0\% purity, Synth\textsuperscript{\textregistered}), with final concentrations of \qty{0.01}{\mol\per\liter} and \qty{5}{\mol\per\liter}, respectively. 

The Er$^{3+}$ and Yb$^{3+}$ dopants were introduced via aqueous solutions of erbium and ytterbium nitrates, which were obtained by dissolving the respective rare-earth oxides (RE$_2$O$_3$, RE = Er, Yb) in a slight excess of nitric acid. The acid excess was evaporated until the solution reached a pH of 4, after which the volume was adjusted to achieve a final concentration of \qty{0.1}{\mol\per\liter}. The dopant concentrations of Er$^{3+}$ and Yb$^{3+}$ were fixed at 0.5 mol\% and 1.5 mol\%, respectively, relative to the molar concentration of Y$^{3+}$. 

The thermolysis reaction was conducted in a sealed vessel at \qty{80}{\degreeCelsius} for 2 hours, allowing for the precipitation of the co-doped precursor nanoparticles. The resulting precipitate was separated by centrifugation at 4000 rpm, washed five times with distilled water, and subsequently dried at \qty{70}{\degreeCelsius} for 6 hours.
The final Er$^{3+}$, Yb$^{3+}$ co-doped Y$_2$O$_3$ nanoparticles were obtained by annealing the Y(OH)CO$_3\cdot$nH$_2$O precursor in air at \qty{900}{\degreeCelsius} for 2 hours, using a controlled heating rate of \qty{5}{\degreeCelsius\per\minute}. X-ray diffraction (XRD) and Transmission Electron Microscopy (TEM) data are shown in Figs. S4a and S4b of the Supporting Information, respectively. he resulting particles had an average diameter of 80 $\pm$ 10 nm. 

\subsection{Experimental Setup}
The dry nanoparticle powder was compacted into a copper sample holder and placed inside a closed-cycle cryostat (Cryostation\textsuperscript{\textregistered} s50 - Montana Instruments), capable of controlling the sample temperature between \qty{4.4}{\kelvin} and \qty{350}{\kelvin}. The cryostat reference temperature has an accuracy of \qty{5}{\milli\kelvin} at \qty{4.4}{\kelvin} and \qty{65}{\milli\kelvin} at \qty{350}{\kelvin}, according to the manufacturer. 

Excitation was performed using a femtosecond laser source (Chameleon\textsuperscript{\textregistered} Ultra II - Coherent) operating at \qty{980}{\nano\meter} with an \qty{80}{\mega\hertz} repetition rate and a spectral width of approximately \qty{10}{\nano\meter}. A \qty{35}{\milli\meter} focal length lens was placed inside the cryostat for both excitation and collection of the luminescence signal in reflection geometry (see supplementary Fig. S5). The excitation beam had a Gaussian profile, with an estimated focal volume of \qty{2e-11}{\cubic\centi\meter}. 

A beam splitter was used to separate the emission from the excitation light. The collected luminescence was directed into a spectrometer (Acton SP2300 - Princeton Instruments), coupled to a CCD camera (Pixis 100F - Princeton Instruments). A 1800 grooves/mm diffraction grating enabled spectral resolution of individual Stark lines. The integration time for all spectra was \qty{60}{\second}.

\begin{acknowledgement}
L. de S. Menezes and T. Possmayer acknowledge the support from the Center for Nanoscience (CeNS), Ludwig Maximilians-Universit\"at M\"unchen, Germany, and the Bavarian program Enabling Quantum Communication and Imaging Applications (EQAP). L. F. dos Santos acknowledges FAPESP (grant numbers: 2020/04157-5 and 2023/03092-5). R. R. Gonçalves acknowledges CNPq (grant number 303110/2019–8 and 306191/2023-7), FAPESP (grant number 2020/05319-9 and 2017/11301-2) and INCT-INFo for financial support. A. M. Amaral and A. R. Pessoa acknowledge the financial support by CAPES and INCT-INFo.
\end{acknowledgement}

\pagebreak

\begin{suppinfo}
Supporting information provides details on sample's characterization and Stark sublevels separation through Voigt profile fitting.
\end{suppinfo}

\vspace{1em}

\textbf{CRediT author statement}: T. POSSMAYER: Investigation, Formal analysis, Writing- Original draft; A. R. PESSOA: Conceptualization, Formal analysis, Writing- Original draft; J. A. O. GALINDO: Formal analysis, Writing- review \& editing; L. F. dos SANTOS: Resources, Writing- Original draft; R. R. GONÇALVES: Resources, Writing- review \& editing; A. M. AMARAL: Supervision, Project Administration, Writing- review \& editing; L. de S. MENEZES: Supervision, Project Administration, Writing- review \& editing.

\onecolumn{
\bibliography{references}}

\pagebreak

\onecolumn{
\begin{figure}[h]
\begin{center}
\includegraphics{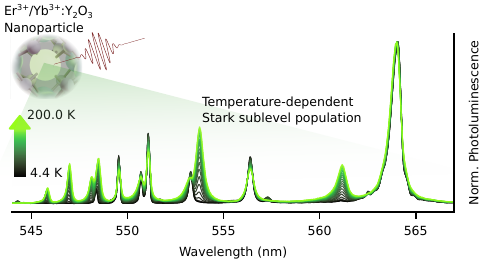}
\caption{TOC Graphic}%
\label{TOC Graphic}
\end{center}
\end{figure}
}

\end{document}


\maketitle

\tableofcontents
\newpage

\section{Stark-resolved spectra}

\begin{figure}[h!] 
	\includegraphics{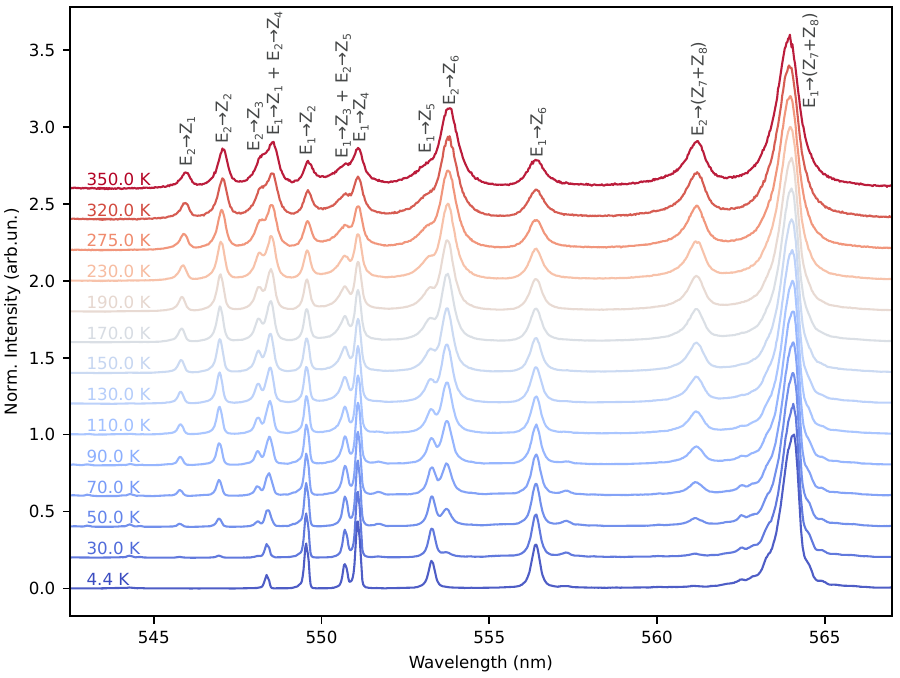}
	\caption{Normalized spectra of the $^4$S$_{3/2}$ $\rightarrow$ $^4$I$_{15/2}$ luminescence band for various temperatures between \qty{4.4}{K} and \qty{350.0}{K} (offset for clarity). The assignments follow the empirical notation as detailed in the main text.}
	\label{fig:simulations}
\end{figure}

The $^4$S$_{3/2}$ $\rightarrow$ $^4$I$_{15/2}$ luminescence band shows multiple sublevels, whose relative individual contributions to the overall emission depend on the temperature due to the different population of the two sublevels E$_1$ and E$_2$. By normalizing the spectra to the brightest peak (E$_1$ $\rightarrow$ (Z$_7$ + Z$_8$)), one can observe that all E$_2$-originating peaks increase with temperature (Fig. \ref{fig:simulations}).

Note that some of the lines - in addition to temperature-induced broadening - experience minor spectral shifts of up to \qty{5}{\per\centi\meter} when going from cryogenic to ambient temperatures, as also reported in other works \cite{Dodson_2024}. Furthermore, certain transitions originating from E$_1$ (e.g., E$_1$ $\rightarrow$ Z$_6$) do not follow the same temperature dependence as the brightest peak, likely caused by a temperature dependence of the radiative decay rate. We attribute both of these changes to the lattice expansion of the surrounding yttria, since the crystal fields impact the resulting Stark wavefunctions \cite{Ciric_2020}.

\FloatBarrier
\FloatBarrier

\section{Integrated intensity}

\begin{figure}[h!] 
	\includegraphics{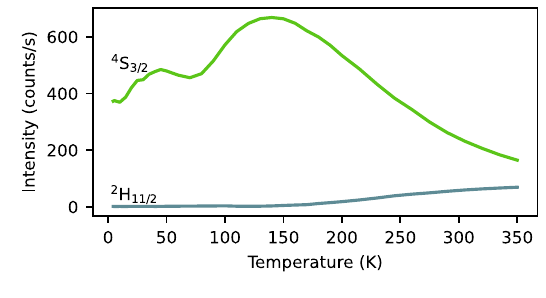}
	\caption{Integrated intensity of the $^2$H$_{11/2}$ $\rightarrow$ $^4$I$_{15/2}$ and $^4$S$_{3/2}$ $\rightarrow$ $^4$I$_{15/2}$ luminescence bands. For each temperature, the spectrum was integrated over $\qty{515}{nm}-\qty{542}{nm}$ for the former and $\qty{542}{nm}-\qty{580}{nm}$ for the latter.}
	\label{fig:overall_intensity}
\end{figure}

Conventional ratiometric luminescence thermometry relies on the integrated intensities of the two manifold transitions $^2$H$_{11/2}$ $\rightarrow$ $^4$I$_{15/2}$ and $^4$S$_{3/2}$ $\rightarrow$ $^4$I$_{15/2}$. Fig. \ref{fig:overall_intensity} shows their temperature-dependent count rates. As expected, the $^2$H$_{11/2}$ manifold has a negligible contribution to the emission spectra below \qty{200}{K}, limiting its usefulness in cryogenic thermometry.

\FloatBarrier
\FloatBarrier

\section{Voigt profile fitting}

To accurately extract intensity ratios from partially overlapping Stark lines, simple spectral integration is insufficient due to finite linewidths and overlap from neighboring transitions. These linewidths arise from both homogeneous and inhomogeneous broadening: the former leads to Lorentzian line shapes, while the latter yields Gaussian components. Accordingly, we fit the emission lines with Voigt profiles, defined as the convolution:

\begin{equation}
V(x) = A_0 \int_{-\infty}^{\infty} G\left(x^{\prime} ; \sigma\right) L\left(x-x^{\prime} ; \gamma\right) \mathrm{d} x^{\prime} \, ,
\end{equation}

\noindent where $G(x ; \sigma)=\frac{1}{\sigma \sqrt{2 \pi}} \exp \left(-\frac{(x-x_0)^2}{2 \sigma^2}\right)$ is the Gaussian component and $L(x ; \gamma)=\frac{\gamma / \pi}{(x-x_0)^2+\gamma^2}$ is the Lorentzian. $A_0$, $x_0$, $\sigma$ and $\gamma$ are the fitting parameters. 

\begin{figure}[h!] 
	\includegraphics{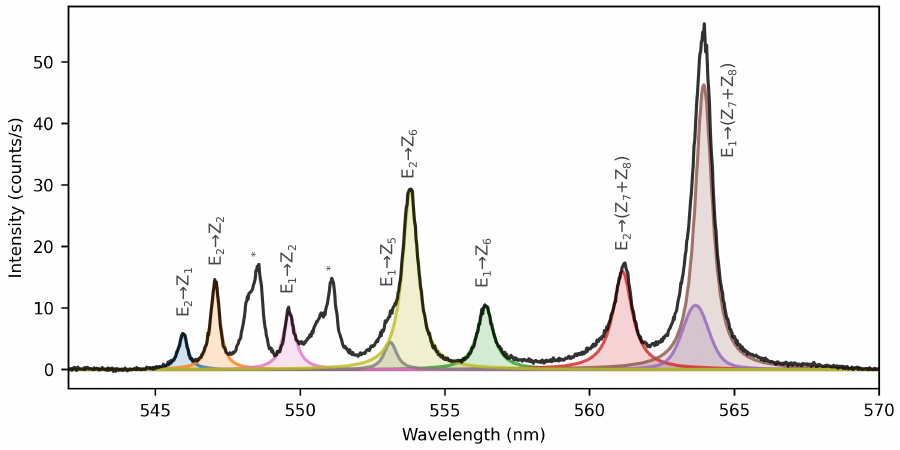}
	\caption{Emission spectrum of the $^4$S$_{3/2}$ $\rightarrow$ $^4$I$_{15/2}$ band at 350 K, with overlaid fitted Voigt profiles. Peaks used in the analysis are labeled; the ones marked with an asterisk (*) are excluded.}
	\label{fig:fitting_voigt}
\end{figure}

Fig. \ref{fig:fitting_voigt} shows the spectrum at \qty{350}{K} with overlaid Voigt fits for each resolvable transition; the two unfitted peaks (marked with an asterisk) contain overlapping contributions from at least three levels, and were therefore excluded from the discussions due to a higher uncertainty in their fit.

\FloatBarrier
\FloatBarrier

\section{Sample characterization}

\begin{figure}[h!] 
	\includegraphics{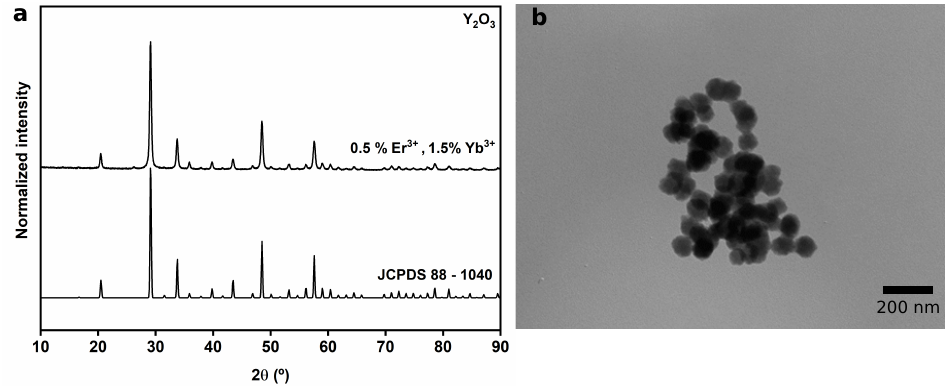}
	\caption{\textbf{a}, X-ray diffractogram of the Y$_2$O$_3$: Yb$^{3+}$ / Er$^{3+}$ nanoparticles used in the experiments, with the corresponding reference spectrum for pure Y$_2$O$_3$ in its cubic phase (JCPDS 88-1040). \textbf{b}, TEM image showing particle morphology.}
	\label{fig:sample_charac}
\end{figure}

The synthesized yttria nanoparticles doped with Yb$^{3+}$ / Er$^{3+}$ were characterized via X-ray diffraction (XRD). Measurements were conducted on a Siemens Bruker D5005 diffractometer operating with CuK$\alpha$ radiation ($\lambda = $ \qty{1.5418}{\angstrom}) and equipped with a graphite monochromator. The diffractogram was recorded with a step of \qty{0.02}{\degree\per\second} in the $2\theta$ range of \qtyrange{10}{90}{\degree\per\second} (Fig. \ref{fig:sample_charac}a). The pattern shows good agreement with the reference spectrum for undoped cubic-phase yttria, confirming that the synthesized material retains a predominantly cubic crystalline structure.

Their morphology was examined under a JEOL JEM-100CX II Transmission Electron Microscope (TEM) operating at 100 kV (Fig. \ref{fig:sample_charac}b). Ethanol was used to disperse the particles, allowing to measure their sizes. Average extracted diameters were $ 80 \pm \qty{10}{nm}$.

\FloatBarrier
\FloatBarrier

\section{Experimental setup}

\begin{figure}[h!] 
	\includegraphics{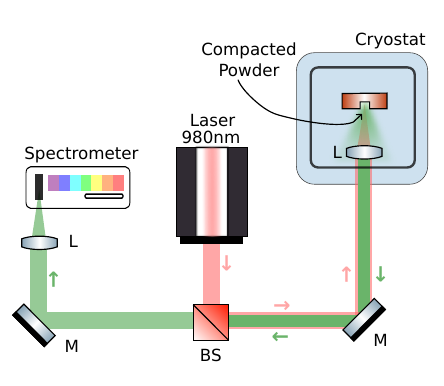}
	\caption{Schematic of the experimental setup for thermometric measurements. L: lens, M: mirror, BS: beam splitter.}
	\label{fig:setup}
\end{figure}

To characterize the sample at cryogenic temperatures, it was mounted in a liquid helium cryostat. Illumination was performed in reflection geometry (Fig. \ref{fig:setup}), using a tunable Ti:Sapphire laser emitting \qty{140}{fs} pulses at a central wavelength of \qty{980}{nm}.

\FloatBarrier
\FloatBarrier

\bibliography{references}